\documentclass[aps,graphicx,showpacs]{revtex4}
\usepackage{graphicx}

\def\ket{\rangle}
\def\<{\langle}
\def\>{\rangle}

\begin{document}
\title{An efficient quantum secret sharing scheme
with Einstein-Podolsky-Rosen Pairs}
\author{Fu-Guo Deng$^{1,2,4}$, Gui Lu Long$^{2,3}$\footnote{gllong@tsinghua.edu.cn (G.L.Long)
\\fgdeng@bnu.edu.cn (F.G.Deng)}
 and Hong-Yu Zhou$^{1,4}$ }
\affiliation{ $^1$ The Key Laboratory of Beam Technology and
Material Modification of Ministry of Education, and Institute of
Low Energy Nuclear Physics, Beijing Normal University, Beijing
100875, China\\
$^2$ Key Laboratory For Quantum Information and Measurements of
Ministry of Education, and Department of Physics, Tsinghua
University,
Beijing 100084, China\\
$^3$ Key Laboratory for Atomic and Molecular Nanosciences,
Tsinghua University, Beijing 100084, China\\
$^4$  Beijing Radiation Center, Beijing 100875,  China}
\date{\today }
\date{\today }

\begin{abstract}
An efficient quantum secret sharing scheme is proposed.  In this
scheme, the particles in an entangled pair group form two particle
sequences. One sequence is sent to Bob and the other is sent to
Charlie after rearranging the particle orders. Bob and Charlie
make coding unitary operations and send the particles back. Alice
makes Bell-basis measurement to read their coding operations.
\end{abstract}

 \maketitle


Secret sharing is a classical cryptographic scheme in which one
use, Alice, can split her message into two parts between two
agents, Bob and Charlie respectively
\cite{Blakley,Blakley2,Blakley3}. It is assumed that Bob and
Charlie can read out the message only when they act in concert.
The combination of quantum mechanics with information theory has
provided us with new tools for many tasks. One mature application
of quantum information is quantum key distribution (QKD) with
which two remote parties of communication can create a private key
unconditionally securely\cite{BB84,gisin}. The quantum version of
secret sharing is  quantum secret sharing(QSS). In quantum secret
sharing, both classical and quantum information can be shared
while the latter is called quantum state sharing \cite{qstates}
which has no classical counterpart. An original QSS scheme that
shares classical information was proposed by Hillery, Bu\v{z}ek
and Berthiaume \cite{HBB99} in 1999, which is called HBB99
hereafter. There have been many theoretical development in this
subject \cite{HBB99,KKI,others,others2,others3,others4,others5,
others6,others7,others8,others9,others10,others11,others12,others13,others14}
and it has been also studied experimentally\cite{TZG}. One
advantage of quantum mechanics is that distribution of the message
can be done securely on-site.

In QSS, Alice has to protect the quantum information from
eavesdropping  by the dishonest one among Bob and Charlie and the
other malicious eavesdroppers. We call the eavesdropper as Bob*.
In the HBB99 scheme, the secret sharing is accomplished by using
three-photon entangled Greenberger-Horne-Zeilinger (GHZ)
states\cite{GHZ}. Each party holds a particle from a GHZ state,
and each participant chooses randomly  from the
$x$-measuring-basis($x$-MB) and the $y$-MB to measure the particle
respectively, a situation similar to the Bennett-Brassard-Mermin
1992 (BBM92) QKD scheme\cite{BBM92}. Karlsson, Koashi and Imoto
put forward a QSS scheme \cite{KKI} with two-photon
polarization-entangled state. In these protocols, the participants
choose randomly one from two measuring-basis to measure the
polarization of the states, and their intrinsic efficiency, the
number of valid particles to the number of total particles, is
50\% because half of the instances are discarded. In addition, in
each accepted round of communication, one bit of information is
shared between the participants. During this process, four bits of
classical communications are required: two bits of information
about the measuring-basis of the two participants and another two
bits of information about the measured results.

In this Letter, we present a QSS scheme with
Einstein-Podolsky-Rosen (EPR) pairs. Our protocols employs the
dense coding \cite{BW} and an order-rearrangement idea. The basic
idea of order-rearrangement is that Alice mixes up the correct
correlation of EPR pairs so that Bob* does not know which two
particles are the particles in an EPR pair and he can not perform
Bell-basis measurement to steal the secret information. Later
Alice restores the correct correspondence of particles and obtains
the result with Bell-basis measurement. In dense coding, a single
qubit can communicate two bits of information by taking advantage
of quantum entanglement. The proposed QSS scheme has three
advantages. First, the intrinsic efficiency is nearly 100\%, all
the EPR pairs except those used for eavesdropping check are
retained for secret sharing. Secondly, inherited from dense
coding\cite{BW}, each EPR pair carries two bits of information.
Thirdly, the classical information exchanged is reduced largely.

An EPR pair is in one of the four Bell states shown as follows:
\begin{eqnarray}
\left\vert \psi ^{-}\right\rangle &=&\frac{1}{\sqrt{2}}(\left\vert
0\right\rangle _{B}\left\vert 1\right\rangle _{C}-\left\vert
1\right\rangle _{B}\left\vert 0\right\rangle _{C})={1\over
\sqrt{2}}(|-x\ket_B|+x\ket_C-|+x\ket_B|-x\ket_C)  \label{EPR1}
\\
\left\vert \psi ^{+}\right\rangle &=&\frac{1}{\sqrt{2}}(\left\vert
0\right\rangle _{B}\left\vert 1\right\rangle _{C}+\left\vert
1\right\rangle _{B}\left\vert 0\right\rangle _{C})={1\over
\sqrt{2}}(|+x\ket_B|+x\ket_C-|-x\ket_B|-x\ket_C)  \label{EPR2}
\\
\left\vert \phi ^{-}\right\rangle &=&\frac{1}{\sqrt{2}}(\left\vert
0\right\rangle _{B}\left\vert 0\right\rangle _{C}-\left\vert
1\right\rangle _{B}\left\vert 1\right\rangle _{C})={1\over
\sqrt{2}}(|-x\ket_B|+x\ket_C+|+x\ket_B|-x\ket_C)  \label{EPR3}
\\
\left\vert \phi ^{+}\right\rangle &=&\frac{1}{\sqrt{2}}(\left\vert
0\right\rangle _{B}\left\vert 0\right\rangle _{C}+\left\vert
1\right\rangle _{B}\left\vert 1\right\rangle _{C})={1\over
\sqrt{2}}(|+x\ket_B|+x\ket_C+|-x\ket_B|-x\ket_C)  \label{EPR4}
\end{eqnarray}%
where $\left\vert 0\right\rangle $ and $\left\vert 1\right\rangle
$\ are the
eigenvectors of  the Pauli operators $%
\sigma _{z}$:
\begin{eqnarray}
\sigma_z|0\ket=|0\ket,\;\;\sigma_z|1\ket=-|1\ket,
\end{eqnarray}
and $|+x\ket$ and $|-x\ket$ are the eigenvectors of the $\sigma_x$
Pauli operator:
\begin{eqnarray}
\sigma_x|+x\ket=|+x\ket,\;\;\sigma_z|-x\ket=-|-x\ket.
\end{eqnarray}
In the Bell states, the results of the $\sigma_z$ or $\sigma_x$
measurement on the two  particles are correlated that can be seen
directly form Eqs.(\ref{EPR1})-(\ref{EPR4}).

An EPR pair can carry two bits of classical information, using the
dense coding operations,
\begin{eqnarray}
U_{0}&&=I=\left\vert 0\right\rangle \left\langle 0\right\vert
+\left\vert 1\right\rangle \left\langle 1\right\vert , \label{eO0}
\\
U_{1}&&=\sigma _{z}=\left\vert 0\right\rangle \left\langle
0\right\vert -\left\vert 1\right\rangle \left\langle 1\right\vert
,  \label{eO1}
\\
U_{2}&&=\sigma _{x}=\left\vert 1\right\rangle \left\langle
0\right\vert +\left\vert 0\right\rangle \left\langle 1\right\vert
,  \label{eO2}
\\
U_{3}&&=i\sigma _{y}=\left\vert 0\right\rangle \left\langle
1\right\vert -\left\vert 1\right\rangle \left\langle 0\right\vert.
\label{eO3}
\end{eqnarray}%
The following naive QSS protocol is not secure: Alice sends the
two particles of an EPR pair to Bob and Charlie respectively. Bob
and Charlie then make an coding operation using one of the
operations in Eqs. (\ref{eO0}),(\ref{eO1}),(\ref{eO2}),(\ref{eO3})
and return the particles to Alice. By making a Bell-basis
measurement, Alice knows the combined operation of Bob and Charlie
which is the shared key. As Bell-basis states are orthogonal
states, Bob* can use the intercept-resend attack to get the secret
information without being detected, as pointed out in
Ref.\cite{KKI}.

 To guard the secret information from eavesdropping,
 one method is not allowing Bob* to acquire simultaneously
both particles of an EPR pair. Because the information is encoded
in the EPR pair, it can only be read by a Bell-basis measurement
if the EPR pair is randomly in one of the four Bell states
\cite{DL}. In the same spirit, if the correct correspondence of
particles in EPR pairs are mixed up, an outsider can not know
which two particles are in the same EPR pairs.  What he can do is
just to make a guess of this correspondence. Inevitably, he will
cause significant errors in the data if he tries to eavesdrop.
This property has been exploited for QKD in Ref. \cite{DL,LL,DLL}.
In the present work, this property is exploited for quantum secret
sharing. Suppose Alice sends the EPR pairs in group of four each
time, seen in Fig.1. In each group, Alice takes one particle from
each pair to form a particle sequence and sends the sequence in
its original order through the AB channel, a channel that connects
Alice and Bob, to Bob. The remaining four particles also form a
particle sequence, and Alice makes an order rearrangement of the
particles and then sends them through the AC channel, a channel
connects Alice and Charlie, to Charlie. Without knowing the
correct correspondence of particles in the AB channel and AC
channel, an eavesdropper can not make orthogonal measurement to
obtain the  deterministic information about the state. Alice has
many different ways to rearrange the particle order in a sequence.
For a group of four objects, there are altogether $4!=24$
permutation operations. Any one of them can be used for the order
rearrangement. Depicted in Fig.\ref{f2} is an example of four such
rearrangement operations. Operation $E_0=I$ is the identity
operation and  no change is made to the particle order, thus the
particle in the AB channel and the corresponding particle in the
AC channel are the two particles in the same EPR pair. If the
rearrangement operation is $E_1$, Alice exchanges the order of
particles 1 and 2, and particles 3 and 4 in the AC channel, and
hence particles 1, 2, 3, 4 in the AB channel and particles 2,1,
4,3 in the AC channel form EPR pairs respectively. Without knowing
the correct particle correspondence, Bob*'s Bell-basis measurement
will obtain no useful information and collapses the quantum
correlation in the genuine EPR pair. This leaves his mark in the
result, and the participants can find Bob* by checking a subset of
results of their measurements.

Now we first give the details of the QSS scheme. For simplicity we
fix the number of EPR pairs in each group to four, and the number
of rearrangement operations is also restricted to four. The
security analysis will be given below. For simplicity we first
assume ideal conditions, perfect detection efficiency and
noiseless channel.

(1) Alice prepares a sequence of EPR pairs randomly in one of the
four Bell-basis states. Alice makes a record of the EPR pair
states. These EPR pairs are divided into groups, each group has
four EPR pairs.

(2) Alice takes each particle from an EPR pair in a group and
sends these four particles in its original order through the AB
channel to Bob. Alice makes an order rearrangement operation to
the remaining four particles using randomly one of four operations
shown in Fig.\ref{f2}, and then sends them through the AC channel
to Charlie.

(3)  After receiving the particles, Bob and Charlie, randomly and
independently  chooses one of the following two modes: with a
small probability $p$ the checking mode, or  with a large
probability $1-p$ the coding mode. If Bob or Charlie chooses the
checking mode, then he goes to step (3a), otherwise he chooses the
coding mode and continues to step (3b).

(3a) When Bob or Charlie chooses the checking mode, he measures
his particle randomly in the the $\sigma_z$ basis  or the
$\sigma_x$ basis , and publishes the position of the measured
particle, but delay the publication of the measured result,
measured quantity ($\sigma_x$ or $\sigma_z$) until he receives
instruction from Alice when she receives both particles of the
same EPR pair.

(3b) When they choose the coding mode, Bob or Charlie performs an
coding operation. The coding operation is one of the four unitary
operations given Eqs.
(\ref{eO0}),(\ref{eO1}),(\ref{eO2}),(\ref{eO3}). These operations
correspond  to the bit values 00, 01, 10 and 11, respectively.
After the coding operations, they return the particles back to
Alice.

(4) After receiving the particles from Bob and Charlie, Alice does
the following actions. For those particles that have not been
chosen for eavesdropping check, Alice first undoes the previous
order rearrangement operation to recover the correct EPR pair
correspondence. Alice then makes a Bell-basis measurement for each
pair, so that she knows the state after the combined operations of
Bob and Charlie. This combined operation of Bob and Charlie are
used as the shared secret key.  When Bob and Charlie work
together, they will know this shared key. The product of
operations of the four coding operations is given in Table
\ref{t1}. The minus sign in this table does not have any physical
effect and is neglected.  From this table, we see that the product
of the two coding operation has a very simple rule in the binary
numbers they represent. The result is the bitwise modulo 2 sum of
the binary numbers of Bob and Charlie, i.e., $K_{A}=K_{B}\oplus
K_{c}$ where $K_A$, $K_B$ and $K_C$ are the binary keys of Alice,
Bob and Charlie respectively.

For those pairs whose partner particle have been chosen by
Bob(Charlie) for eavesdropping check, Alice makes randomly the
$\sigma_z$ or $\sigma_x$ measurement on the particle returned from
Bob(Charlie), and using the same measuring-basis to measure the
particles returned from Charlie(Bob). At this moment, Alice asks
Bob(Charlie) to publish the measured quantity, and the measurement
value of the particles, and asks Charlie(Bob)to publish the coding
operations performed on the partner particles. In half of these
instances, Alice chooses the same measuring-basis as Bob(Charlie).
From the expressions for Bell-basis states, the results of
measurement are correlated. With this knowledge, Alice checks the
consistence of the results to find eavesdropping. There are also a
small portion in which the same EPR pair are chosen simultaneously
by Bob and Charlie for eavesdropping check. In half the case, Bob
and Charlie choose the same measuring-basis, and these instances
can also be used for eavesdropping check. If the error rate is
high, they conclude the QSS process as insecure and abort the
process. If there is no error in the eavesdropping check, they
conclude the QSS as secure and continues to step 5).

 (5)Alice publishes the order rearrangement operation for each
 group, and those particles that have been chosen for
 eavesdropping check.
With these information, Bob and Charlie get the correct
correspondence of their particles. The production of their coding
operation is the key Alice wants Bob and Charlie to share.

For preventing a dishonest agent from eavesdropping with a fake
signal, Alice inserts some decoy photons in some of the EPR
groups. That is, Alice replaces some of the particles in EPR pairs
with decoy photons which can be produced by means that she
measures one particle in an EPR pair with choosing $\sigma_z$ or
$\sigma_x$ randomly.  If the dishonest agent intercepts these
decoy photons in the EPR-pair groups and resends a fake signal,
his action will introduce inevitably errors in the results of the
decoy photons which is chosen by the other agent and measured with
the two measuring bases, $\sigma_z$ and $\sigma_x$, for
eavesdropping check, same as BB84 QKD protocol \cite{BB84}.

In this scheme, almost all of the instances except those chosen
for eavesdropping can be used for quantum secret sharing. This
scheme is thus efficient. Furthermore,  each successful EPR pair
carries two qubits of information in this scheme. Moreover, the
three parties in the communication use only a small amount of
classical communication. Alice needs to publish only the order
rearrangement operation for each group, and Bob and Charlie need
only to publish the following information: positions,
measuring-basis and measured results and the coding operations of
the particles that have been chosen for eavesdropping checking
analysis.

First we explain the working mechanism of the checking mode. For
instance, when Bob chooses the checking mode for a particle from
an EPR pair in state
$|\psi^-\ket=\frac{1}{\sqrt{2}}(|0\ket_B|1\ket_C-|1\ket_B|0\ket_C)$,
and he makes a $\sigma_z$ measurement to this particle and obtains
$|0\ket_B$, and as a result, the corresponding particle $C$ that
is in Charlie's site collapses into state $|1\ket_C$.  Charlie
does not know Bob has chosen this pair for eavesdropping check,
and he chooses for this particle to perform the $\sigma_z$
measurement with a small probability, and obtains, say $|1\ket_C$
if no eavesdropping exists. Charlie publishes the position of the
particle he measures, and delays the publication of the
measuring-basis and the measured result until he receives
instruction from Alice. With a large probability, Charlie will
choose the coding mode and performs a coding operation on it, say
$U_3$. This changes the state of his particle from $|1\ket_C$ to
$|0\ket_C$, and returns the particle back to Alice. As Alice knows
that this particle C is the partner particle of particle B which
was measured by Bob, Alice measures $\sigma_z$ for particle C. If
no eavesdropping exists, she will obtain $|0\ket_C$. At this
point, Alice asks Bob to publish his measuring-basis and the
measured result, and asks Charlie to publish his coding operation.
With these knowledge, Alice can check the consistence of the
measured results.

This eavesdropping check guard  against the intercept-resend
attack. For instance, suppose Eve can intercept all particles sent
to Bob and Charlie, and stores them for a while. At the same time,
he sends fake particles to Bob and Charlie respectively. Without
knowing the particles are fake, Bob and Charlie encode the unitary
operations respectively on their particles and return them back.
Bob* intercepts them again and measures their states and steals
the operations of Bob and Charlie. He then performs these
operations on the genuine particles he stores previously and sends
them back to Alice. A collapse measurement in the checking mode
exposes Eve's interception.

The QSS protocol is equivalent to a modified BBM92 QKD protocol
for those chosen for the eavesdropping check. For these instances
where Charlie(Bob) chooses randomly to measure in the $\sigma_x$
or the $\sigma_z$ basis, Alice is in effect doing a BBM92 QKD
process with Charlie(Bob) with an eavesdropper Eve present. If Bob
performs honestly, then there is no error rate in these checking
instances. If Bob is dishonest, the error rate will be as high as
25\%, just like that in the BBM92 QKD protocol. The proof of
security for BBM92 QKD in ideal condition is given in
Ref.\cite{IRV} and that with practical conditions was given in detail in Ref.%
\cite{WZY}. Hence the present QSS protocol is secure. The first
eavesdropping check thus ensures that the particles arrive Bob and
Charlie's sites securely.

When Bob* tries to steal the information by making Bell-basis
measurement  on pairs of particles returned by Bob and Charlie.
Bob* has to make a guess of the correspondence of the particles.
But he will cause a large error rate. Suppose that Alice uses only
four permutations to reshuffle the order of particles in each
group of four EPR pairs, Bob* has only $1/4$ probability to make
the right guess. For those wrongly chosen pairs,  the two
particles he measures is uncorrelated, say particle B from the
first EPR pair and particle C from the second EPR pair is
mistreated by Bob* as an EPR pair, then the density operator is
\begin{equation}
\rho _{B_{1}C_{2}}=\overline{\rho }_{B_{1}}\otimes \overline{\rho }%
_{C_{2}}=\left(
\begin{array}{cccc}
1/4 & 0 & 0 & 0 \\
0 & 1/4 & 0 & 0 \\
0 & 0 & 1/4 & 0 \\
0 & 0 & 0 & 1/4%
\end{array}%
\right)   \label{matrix}
\end{equation}%
where $\overline{\rho }_{B_{1}}=Tr_{C_{1}}(\rho _{B_{1}C_{1}})$ and $%
\overline{\rho }_{C_{2}}=Tr_{B_{2}}(\rho _{B_{2}C_{2}})$ are the
reduced density matrices of particle B$_{1}$ and particle C$_{2}$,
respectively. When $\rho _{B_{1}C_{2}}$ is measured in the
Bell-basis, the result can be any of the 4 Bell-basis states with
25\% probability each. Thus Bob* will introduce 3/4 $\times $
3/4=56.25\% error rate in the results. They can detect Bob* easily
by checking a sufficiently large subset of results randomly
chosen. If Alice allows more number of different permutations, the
error rate Bob* introduces in this type of attack is even greater.

In Refs.\cite{fggnp,ssz,sg}, the optimal individual attack done by
Bob* can be represented by an unitary operation on the travelling
particle sent to, say to  Charlie, and then back to Alice with an
auxiliary system whose initial state is $\left\vert 0\right\rangle
$, i.e.,
\begin{eqnarray}
U_{TB}\left\vert \xi \right\rangle \left\vert 0\right\rangle
&&=\left\vert \xi \right\rangle \left\vert 0\right\rangle
\label{ut1} \\
U_{TB}\left\vert \overline{\xi }\right\rangle \left\vert
0\right\rangle &&=\cos \phi \left\vert \overline{\xi
}\right\rangle \left\vert 0\right\rangle +\sin \phi \left\vert \xi
\right\rangle \left\vert 1\right\rangle \label{ut2}
\end{eqnarray}%
where $\left\vert \xi \right\rangle $ and $\left\vert \overline{\xi }%
\right\rangle $\ are the two eigenvectors of a two-level
operator\cite{fggnp,ssz}, and $\phi \in \lbrack 0,\pi /4]$
characterizes the strength of Bob*'s attack\cite{sg}. This
eavesdropping does not violate the non-cloning
theorem\cite{nocloning} as Bob* just exchanges the state of the
particle of Charlie with the auxiliary state conditionally. So the
information ($I_{B}$) that Bob* can steal from the particle coded
by Charlie is less than twice the information ($I_{0}$) carried by
the particle sent from Alice to Charlie. Although the particle for
Charlie carries 1 bit of quantum information and the four quantum
operations can send 2 bits of information, Bob* will at best get
the two bits of information if he does the eavesdropping. Then
$I_{B}\leq 2I_{0}$.

We can calculate the relations between the information Bob* can
obtain and the error rate introduced by her disturbance. For Alice
and Charlie, the action of Bob*'s eavesdropping
will introduce an error rate%
\begin{equation}
\varepsilon =P\overline{_{\xi }}\sin ^{2}\phi,   \label{e1}
\end{equation}%
where $P\overline{_{\xi }}$ is the probability that the quantum
signal is in state $\left\vert \overline{\xi }\right\rangle $. Let
us suppose that the entangled state between Alice and Charlie is
$\left\vert \psi ^{-}\right\rangle $. In this case, $\left\vert
\xi \right\rangle =\left\vert 0\right\rangle $ and $\left\vert
\overline{\xi }\right\rangle
=\left\vert 1\right\rangle $ are the eigenvectors of $\sigma _{z}$, $P%
\overline{_{\xi }}=\frac{1}{2}$. The effect of Bob*'s
eavesdropping is
\begin{eqnarray}
U_{TB}\left\vert 0\right\rangle \left\vert 0\right\rangle
&&=\left\vert 0\right\rangle \left\vert 0\right\rangle \equiv
\left\vert 00\right\rangle, \label{ut3}\\
 U_{TB}\left\vert 1\right\rangle \left\vert 0\right\rangle &&=\cos
\phi \left\vert 1\right\rangle \left\vert 0\right\rangle +\sin
\phi \left\vert 0\right\rangle \left\vert 1\right\rangle =\cos
\phi \left\vert 10\right\rangle +\sin \phi \left\vert
01\right\rangle,   \label{ut4} \\
 \varepsilon &&=\frac{1}{2}\sin
^{2}\phi .  \label{e2}
\end{eqnarray}%
After the eavesdropping, the state of the system composed of the
particles belong to Alice and Charlie, and the auxiliary system is
\begin{eqnarray}
\left\vert S\right\rangle  &=&\frac{1}{\sqrt{2}}(\left\vert
0\right\rangle _{A}U_{TB}\left\vert 1\right\rangle _{C}\left\vert
0\right\rangle _{P}-\left\vert 1\right\rangle _{A}U_{TB}\left\vert
0\right\rangle
_{C}\left\vert 0\right\rangle _{P})  \label{system} \\
&=&\frac{1}{\sqrt{2}}[\left\vert 0\right\rangle _{A}(\cos \phi
\left\vert 1\right\rangle _{C}\left\vert 0\right\rangle _{P}+\sin
\phi \left\vert 0\right\rangle _{C}\left\vert 1\right\rangle
_{P})-\left\vert 1\right\rangle _{A}\left\vert 0\right\rangle
_{C}\left\vert 0\right\rangle _{P}],  \nonumber
\end{eqnarray}%
where the subscript $P$ represents the auxiliary system.

For the sub-system that composed of Alice's particle and the
auxiliary system, its density matrix $\rho $\ is obtained by
tracing out the freedom of Charlie's particle, i.e.,
\begin{eqnarray*}
\rho  &=&\frac{1}{2}\cos ^{2}\phi \left\vert 0
\right\rangle_{A}\left\vert 0\right\rangle _{P\;A}
\left\langle 0\right\vert _{P}\left\langle 0\right\vert +%
\frac{1}{2}\sin ^{2}\phi \left\vert 0\right\rangle _{A}\left\vert
1\right\rangle_{P\;A}\left\langle 0\right\vert _{P}\left\langle 1\right\vert +%
\frac{1}{2}\left\vert 1\right\rangle_{A}\left\vert 0
\right\rangle_{P\;A}\left\langle 1\right\vert _{P}\left\langle 0\right\vert  \\
&&-\frac{1}{2}\sin \phi \left\vert 0\right\rangle _{A}\left\vert
1\right\rangle _{P\;A}\left\langle 1\right\vert _{P}\left\langle 0\right\vert -%
\frac{1}{2}\sin \phi \left\vert 1\right\rangle _{A}\left\vert
0\right\rangle _{P\;A}\left\langle 0\right\vert _{P}\left\langle
1\right\vert,
\end{eqnarray*}%
which can be rewritten in the basis $\left\{ \left\vert
00\right\rangle ,\left\vert 11\right\rangle ,\left\vert
01\right\rangle ,\left\vert 10\right\rangle \right\} $ as
\begin{eqnarray}
\rho =\left(
\begin{array}{cccc}
\frac{1}{2}\cos ^{2}\phi  & 0 & 0 & 0 \\
0 & 0 & 0 & 0 \\
0 & 0 & \frac{1}{2}\sin ^{2}\phi  & -\frac{1}{2}\sin \phi  \\
0 & 0 & -\frac{1}{2}\sin \phi  & \frac{1}{2}%
\end{array}%
\right).
\end{eqnarray}
The information $I_{B}$ that Bob* can extract from the sub-system
is just equal to the Von Neumann entropy \cite{DLL}, i.e.,
\begin{equation}
I_{B}=\sum\limits_{i=0}^{3}-\lambda _{i}\log _{2}\lambda _{i},
\label{information1}
\end{equation}%
where $\lambda _{i}$ (i=0, 1, 2, 3) are the eigenvalues of $\rho $, which are $%
\lambda _{0,1}=0$, $\lambda _{2}=\frac{1}{2}\cos ^{2}\phi $ and
$\lambda
_{3}=\frac{1}{2}+\frac{1}{2}\sin ^{2}\phi $, respectively. Then%
\begin{eqnarray*}
I_{B} &=&-\frac{1}{2}\cos ^{2}\phi \log _{2}(\frac{1}{2}\cos ^{2}\phi )-(%
\frac{1}{2}+\frac{1}{2}\sin ^{2}\phi )\log
_{2}(\frac{1}{2}+\frac{1}{2}\sin
^{2}\phi ) \\
&=&-(\frac{1}{2}-\varepsilon )\log _{2}(\frac{1}{2}-\varepsilon )-\frac{%
(1+\varepsilon )}{2}\log _{2}\frac{(1+\varepsilon )}{2}.
\end{eqnarray*}%
When $\varepsilon =0.25$, $I_{B}=1$ which is the maximal
information that Bob* can obtain from his eavesdropping which is
the one in the the intercept-resend eavesdropping strategy. In
other cases, $I_{0}<1=I_{AC}$, where $I_{AC}$ is the mutual
information between Alice and Charlie. Then this QSS is secure.

Under practical conditions, the detection efficiency is not equal
to 1, and the quantum channel is noisy. Then error corrections and
post-processing have to be used. Accordingly, the step 4) of this
QSS should change to accommodate some errors in the process: if
the error rate is below certain threshold $\epsilon$, the QSS
process is then concluded as secure. If the error rate is higher
than the threshold, the QSS process is concluded as insecure, and
the process halts. The exact value of $\epsilon$ depends on the
way the quantum correction is done. For instance for BB84 QKD
protocol, the $\epsilon$ is 7\% for  the error-correcting and
post-processing methods  proposed in Refs.\cite{mayers,biham}, and
11\% by the methods proposed in Refs.\cite{lo-chau,preskill}, and
19\% by the Gotteman and Lo method\cite{gottesman}. These details
are important and merit detailed analysis, and will not be studied
in this Letter.

In summary, we introduce a QSS protocol using  the ideas from dense coding%
\cite{BW} and order rearrangement of EPR pairs. In this QSS
scheme, the amount of information carried by each EPR pair is
 two bits for each successful
round of communication. The amount of classical communication
exchanged in the process is also reduced. The scheme is efficient
since all the EPR pairs can be used for quantum secret sharing
except those that  chosen for eavesdropping check.  Like the
KKI\cite{KKI} QSS scheme, this scheme uses EPR pairs, which is
easier to create rather than three-partitite entangled state. With
these two techniques, not only QSS can be done securely with EPR
pairs, but also the capacity is increased to 2 bits for each EPR
pair and the classical information exchanged is reduced to
$\frac{1}{4}$ bit for each useful qubit on the average.

This work was supported by the National Fundamental Research
Program Grant No. 001CB309308, China National Natural Science
Foundation Grant No. 60073009,10325521,10447106, the Hang-Tian
Science Fund, and the SRFDP program of Education Ministry of
China.

\begin{table}\begin{center}\begin{tabular}{c|cccc}\hline
            & \multicolumn{4}{c}{$U_j$}\\ \cline{2-5}
 $U_i$      & $U_0$ & $U_1$ & $U_2$ & $U_3$ \\ \hline
$U_0$ & $U_0$ & $U_1$ & $U_2$ & $U_3$ \\
$U_1$ & $U_1$ & $U_0$ & $-U_3$ & $U_2$ \\
$U_2$ & $U_2$ & $-U_3$ & $U_0$ & $-U_1$ \\
$U_3$ & $U_3$ & $U_2$ & $U_1$ & $U_0$ \\
\hline
\end{tabular} \caption{Table of products of $U_i\times
U_j$. }\label{t1}\end{center}
\end{table}

\begin{figure}
\begin{center}
\caption{ Illustration of the QSS scheme with order rearrangement
apparatus (REAR system). }
\includegraphics[width=12cm,angle=0]{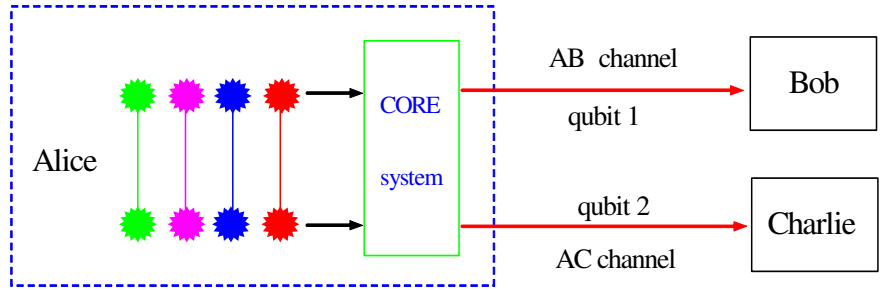} \label{f1}
\end{center}
\end{figure}

\begin{figure}
\begin{center}
\caption{ Four rearrangement operations for EPR pairs. }
\includegraphics[width=12cm,angle=0]{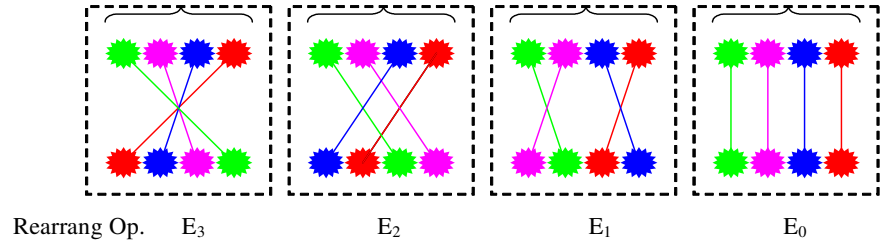} \label{f2}
\end{center}
\end{figure}

\end{document}